# Schottky barrier induced asymmetry in the negative differential resistance response of Nb/NbO$_x$/Pt cross-point devices

*Shimul Kanti Nath[†*], Sanjoy Kumar Nandi[†], Assaad El-Helou[‡], Xinjun Liu[§], Shuai Li[†,#], Thomas Ratcliff[†], Peter E Raad[‡], Robert G Elliman[†*].*

[†]Department of Electronic Materials Engineering, Research School of Physics, The Australian National University, Canberra ACT 2601, Australia

[‡]Department of Mechanical Engineering, Southern Methodist University, Texas, Dallas, USA

[§]Tianjin Key Laboratory of Low Dimensional Materials Physics and Preparation Technology, Faculty of Science, Tianjin University, Tianjin 300354, China

[#]Unité Mixte de Physique, CNRS, Thales, Université Paris-Sud, Université Paris-Saclay, Palaiseau, France.

**ABSTRACT:** The negative differential resistance (NDR) response of Nb/NbO$_x$/Pt cross-point devices is shown to have a polarity dependence due to the effect of the metal/oxide Schottky barriers on the contact resistance. Three distinct responses are observed under opposite polarity testing: bipolar S-type NDR, bipolar snap-back NDR, and combined S-type and snap-back NDR, depending on the stoichiometry of the oxide film and device area. In-situ thermoreflectance imaging is used to show that these NDR responses are associated with strong current localisation, thereby justifying the use of a previously developed two-zone, core shell thermal model of the device. The observed polarity dependent NDR responses, and their dependence on stoichiometry and area are then explained by extending this model to include the effect of the polarity dependent contact resistance. This study provides an improved understanding of the NDR response of metal/oxide/metal structures and informs the engineering of devices for neuromorphic computing and non-volatile memory applications.

**INTRODUCTION**

Two terminal metal-oxide-metal (MOM) devices using transition metal oxides (TMOs) are of increasing interest for next generation non-volatile memory technology and brain-inspired computing[1-4]. Many TMOs exhibit threshold switching or current-controlled negative differential resistance (NDR) response including $NbO_x$[1, 3, 5-7], $TaO_x$[8-10], $VO_x$,[11-12] $TiO_x$[13] and $NiO_x$[14] and MOM devices exploiting this characteristics are of great interest as self-sustained and chaotic oscillator[2, 15], threshold logic device[16-17], trigger comparator[18], small signal amplifier[19-20] and emulator of biological neuronal functionalities[21-22]. In the recent years $NbO_x$ and $VO_x$ have drawn much attention due to their ability to show reliable threshold switching [23] and stable current-controlled negative differential resistance[1-2, 5, 24]. Importantly, the functionality of the device is greatly controlled by their specific current-voltage characteristics (e.g. specific NDR mode) [3, 15, 25], and therefore it is essential to understand the physical origin of the NDR response and its dependencies on materials and device parameters.

In many cases the NDR response is only observed after an initial electroforming process which introduces a filamentary conduction path through the film[23]. The filament is created by the generation, drift and diffusion of oxygen vacancies and can result in competition between resistive and threshold switching, in which case it is necessary to impose constraint on the operating current and/or specific bias direction in order to achieve pure threshold switching[26]. Threshold switching then occurs in a small volume between the residual filament and electrode due to the localised current flow in this region[26]. In other cases, the forming process simply initiates threshold switching, possibly due to reduction of the oxide at the electrode interface produced by local Joule heating. Interestingly, filamentary conduction is expected in materials exhibiting NDR even when there are no material inhomogeneities and already demonstrated in high conductivity $NbO_x$ films[27]. Studies have also shown that several NDR modes including two basic types: continuous 'S-type' and discontinuous 'snap-back' NDR as well as more complex combinations can be observed in $NbO_x$ systems[6, 25]. The physical origin of the S-type NDR is well explained by a thermally induced conductivity changes due to local joule heating[7, 28] and it has been shown that it can, in principle, arise from any electrical conduction mechanism that shows super-linear conductivity change as a function of temperature[29]. In contrast, the origin of snapback NDR was attributed to the insulator-metal transition (IMT) associated with a Mott-Peierls transition observed in $NbO_2$[6]. However this was further questioned in a separate study based on finite-element modelling of $TaO_x$ and $VO_x$ devices that reproduces the snap-back response without recourse to a material specific phase transition[8].

In an attempt to resolve this ensuing debate about the origin of the snap back response and also to explain a diverse range of NDR modes observed in metal-insulator-metal (MIM) devices, we recently developed a core-shell model of NDR and demonstrated its utility to predict novel functionality for emerging electronic and neuromorphic computing applications[25]. This model is based on the fact that the current distribution in metal-insulator-metal (MIM) structure is highly localized after electroforming, with the NDR response dominated by the region of highest current density (core) and the surrounding film (shell) acting as a parallel resistance. The threshold switching response then depends on the relative magnitudes of the maximum negative differential resistance ($R_{NDR}$) of the core and the shell resistances ($R_S$). Specifically, S-type NDR is observed for $R_S > R_{NDR}$ and snap-back characteristic is observed for $R_S < R_{NDR}$, notably, without considering any phase transition. Our model predicts that the threshold switching characteristics depends critically on the current distribution in the device structure and therefore is highly sensitive to the oxide conductivity, device geometry and metal/oxide interface reactions. Indeed, we have verified some of the model predictions by showing a clear transition between smooth S-type NDR and discontinuous snap back response depending on the oxide stoichiometry, thickness and area of the active MOM device[30]. On the other hand, a number of studies have shown that the metal/oxide interfaces can play crucial role in a MOM system by influencing the dominant resistive switching mode, device stability and reproducibility [2, 31]. Therefore, further studies are clearly required to understand the impact of metal/oxide interfaces in determining the NDR response.

In the present work, we demonstrate that the interface resistance originated from the Schottky barriers at the metal/oxide interfaces can significantly influence the threshold switching properties of an electroformed MOM device and can result in distinct NDR modes depending on the operating bias polarity. The observed materials and device dependencies have been analysed by combining in-situ thermo-reflectance imaging and temperature dependent electrical testing. The results are explained by extending the core-shell model of NDR.

**EXPERIMENTAL METHODS**

Cross-point devices with a metal/oxide/metal structure as shown in Figure 1 were fabricated on a thermally oxidized Si (100) wafer with a 300 nm thick oxide layer using a three-mask photolithography process. The bottom electrodes were defined by a lift-off process and consisted of a 5 nm Ti adhesion layer and a 25 nm Pt electrode layer, deposited by e-beam evaporation. These were coated with sub-stoichiometric $NbO_x$ layers by reactive sputter deposition of an Nb target using an Ar/O ambient. For comparison, near-stoichiometric $Nb_2O_5$ film were also deposited by RF-sputtering of an $Nb_2O_5$ target (Ar ambient). Top electrodes were subsequently defined by a second lift-off process and comprised 5 nm Nb and 25 nm Pt (or 75 nm Au for the thermo-reflectance measurements). A final lithography step was then used to remove the $NbO_x$ film from the contact-pads of the bottom electrodes. Details of the deposition conditions and film thicknesses are given in Table-1 (supplementary information). The resulting cross-point structures had dimensions ranging from 2μm x 2μm to 20μm x 20μm.

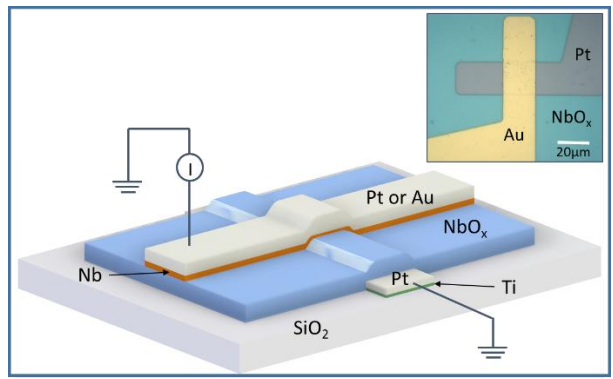

**Figure 1**: Three-dimensional schematic of the cross-point device with corresponding material layers. Inset shows an optical microscopy image of typical Au/Nb/$NbO_x$/Pt cross-point device with 20μm x 20μm active device area.

The $NbO_x$ stoichiometry was determined by Rutherford backscattering spectrometry (RBS) of films deposited onto vitreous carbon and Si substrates and was found to be in the range from

x=1.92±0.03 to x=2.6±0.05. (The relationship between stoichiometry and film resistivity is included in the Supplementary Information). Further analysis by grazing incident-angle X-ray diffraction (GIAXRD) showed that the as-deposited films were amorphous. Ex-situ electrical measurements were performed in air using an Agilent B1500A semiconductor parameter analyser attached to a Signatone probe station (S-1160) and generally consisted of bidirectional quasi-static current sweeps. In-situ measurements performed during thermoreflectance imaging were performed with a Keithley 2410 parameter analyzer and included both DC and pulsed testing. In all cases the bias was applied through the top electrode while the bottom electrode was grounded. Thermal imaging of the devices was performed using the TMX T°Imager® system. The imaging setup uses a camera-based thermoreflectance method [32] to infer the temperature rise from the change in surface reflection. The measured map of the temperature rise ΔT is determined by the use of the relation, $\Delta T = \frac{1}{C_{TR}} \frac{\Delta R}{R}$; where ΔR is the change in reflectance of the activated device surface. $C_{TR}$ is the thermoreflectance coefficient of the top electrode material (Au in the present case) obtained from calibration according to $C_{TR} = \frac{1}{\Delta T} \frac{\Delta R}{R}$ where ΔR is the change in reflectivity of the device surface when subjected to a known temperature rise, $\Delta T$.

**RESULTS AND DISCUSSION**

**Electroforming and NDR characteristics.**

All devices with sub-stoichiometric NbO$_x$ films exhibited negative differential resistance (NDR) following a one-off electroforming step but the forming and NDR characteristics were found to depend on the measurement polarity, as illustrated in Figure 2 for 5 μm NbO$_{1.92}$ devices. During an initial positive bias sweep from 0→8 mA (Figure 2a) the device undergoes an electroforming step characterised by a sudden voltage drop at a current ~6 mA and a permanent change in the low-field device resistance from ~5 kΩ to ~4.5 kΩ (measured at 0.25V). The impact of electroforming is then clearly evident during the reverse current sweep (from 8→0 mA) which shows S-type NDR. Subsequent current sweeps under positive bias continued to show S-type characteristics under both forward and reverse sweeps (similar to that shown by the red line in Figure 2(c)).

Similar electroforming characteristics were observed under negative bias during the forward sweep from 0→10 mA (Figure 2b), with a sudden voltage drop observed at a current of ~7 mA. However, in this case, the reverse current sweep produced a snap-back characteristic, with an abrupt voltage increase as the current was reduced to ~2.2 mA. Subsequent current sweeps under negative bias showed similar snap-back characteristics for both forward and reverse sweeps, but at a lower threshold-current and voltage (similar to that shown by the black line in Figure 2(c)). Once electroformed, devices exhibited characteristics similar to those depicted in Figure 2(c), regardless of the initial electroforming polarity. i.e. Under positive bias they exhibited S-type NDR during forward and reverse current sweeps and under negative bias they exhibited a snap-back response (Further examples of polarity dependent electroforming and subsequent I-V sweeps are included in the Supplementary Information).

The permanent changes produced by electroforming are typically attributed to the creation of a filamentary conduction path in the oxide film and/or local breakdown of the Schottky barriers at the metal/oxide interfaces[27]. From a circuit perspective this represents a parallel conduction path within the device and the total device current is divided between the filamentary conduction path and the surrounding device area based on their relative resistances[30]. Indeed, even though the filamentary region may have a higher conductivity, its resistance can exceed that of the surrounding device due to its smaller effective area.

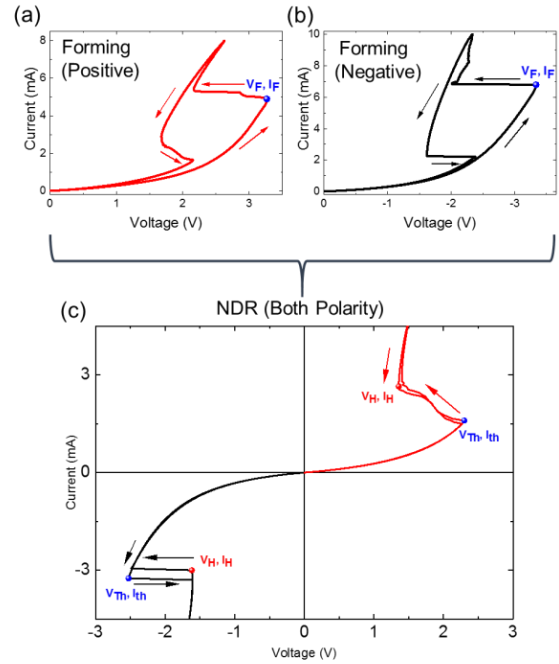

**Figure 2**: Polarity dependent electroforming and CC-NDR characteristics of Pt/Nb/NbO$_{1.92}$/Pt devices: (a) electroforming of a 5 μm × 5 μm cross-point device under positive bias, (b) electroforming of a different 5 μm × 5 μm cross-point device under negative bias; and (c) typical polarity dependent switching characteristics of electroformed devices showing a transition of S-type into snap-back NDR when a negative bias is applied to the top electrode and vice versa irrespective of electroforming polarity.

**In-situ temperature mapping.**

In-situ thermo-reflectance imaging was performed on 10 μm NbO$_{1.99}$ devices to better understand the current distribution during post-forming current sweeps. The results are summarised in Figure 3 and include the in-situ current-voltage (I-V) characteristics and the measured temperature distributions for devices subjected to positive and negative bias. Under positive bias the temperature distribution is highly localised in the filamentary region over the entire range of operating currents and the temperature in this region and that of the surrounding film increase monotonically with increasing current. This clearly demonstrates that the resistance of the filamentary path is lower than that of the surrounding device and that the S-type NDR characteristic is dominated by the temperature-dependent conductivity of the filamentary region.

In contrast, under negative bias, the temperature distribution is near-uniform at currents below the threshold for snap-back and only becomes localised for currents near the threshold value (indicated by the point 'F' in Figure 3(b)). At this point the temperature increases abruptly while that of the surrounding area decreases, consistent with current redistribution and localisation due to the positive feedback created by local Joule heating, as previously reported[30]. This demonstrates that the resistance of the filamentary region is initially comparable to, or greater than that of the surrounding device, but becomes relatively less resistive as the current increases to the threshold value. Such a change implies that the conductivity of the filamentary region increases more rapidly with temperature than that of the surrounding device. For filaments near the centre of the device this may also be facilitated by the non-uniform temperature distribution created by the device geometry (see Figure 3(c)).

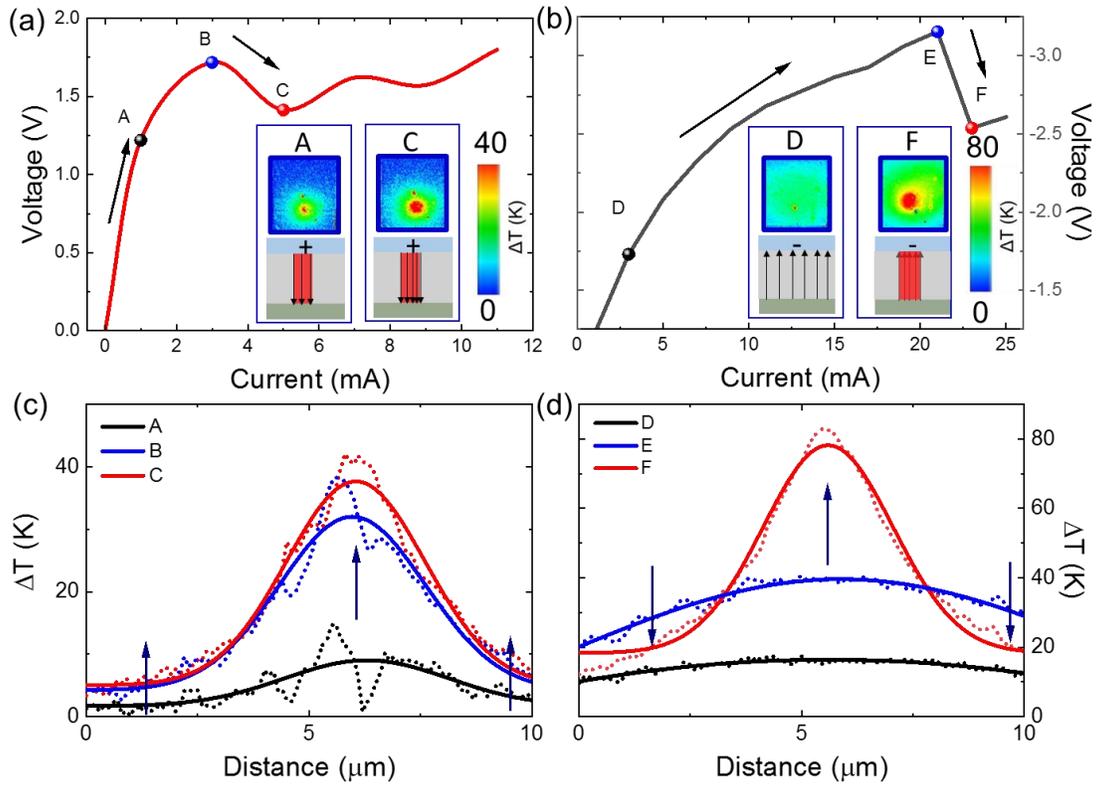

**Figure 3:** (a-b), In-situ I-V characteristics of a 10μm × 10 μm Au (75nm)/Nb(5nm)/NbO$_{1.99}$(~45nm)/Pt (25nm) device showing S-type NDR under positive bias and snapback under negative bias, respectively. Insets show the in-situ thermoreflectance (ΔR/R) maps of the 10μm × 10μm device during corresponding bias polarity, (c) Temperature profile through the filamentary region for different points in the current-voltage curve during positive bias, and (d), Temperature profile of the same device for different points in the current-voltage curve during negative bias. Note: The temperature map and the I-V characteristics shown in Fig. 3 were also obtained for 5μm × 5 μm device devices with lower threshold currents, and a snap back response observed under negative polarity and S-type NDR under positive polarity. But in this case, the filament was observed to form around the edges due to the dominance of edge effect in smaller area cross-point devices with a thick (75nm Au in this case) capping layer[33]. The short-dashed

curves in (c-d) represent the experimental data and the smooth lines are showing Gaussian fits of the corresponding experimental data. A, B and C denote subthreshold, threshold and post-threshold points respectively in the I-V characteristic obtained under positive bias, while D, E and F denote the same for negative bias polarity

**Schottky barriers and polarity dependent conduction.**

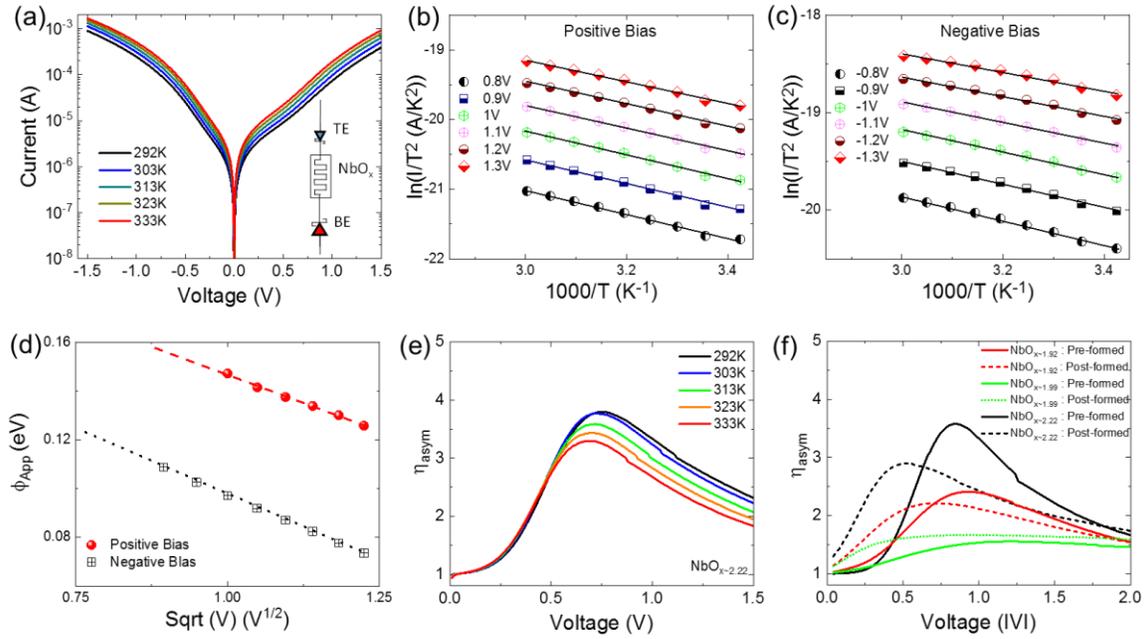

**Figure 4**: (a) Temperature dependent current-voltage characteristics of a typical sub-stoichiometric NbOx device (x=2.22), equivalent representation of the device in terms of electronic circuit element is shown in inset, (b-c) signature plots confirming thermionic emission as the major mechanism for Nb top and Pt bottom electrodes (x = 2.22), (d) extraction of zero bias potential barrier at top and bottom interfaces for a typical sub-stoichiometric NbOx (x= 2.22) device, (e) asymmetry plot of the data presented in Figure 4(a), current asymmetry $\eta_{asym}$ is defined as the ratio of device current under negative bias to that under positive bias ($I_{negative}/I_{positive}$), and

(f) current asymmetry before and after electroforming for devices with different oxide stoichiometry, The dashed lines in Figure 4(f) represent the experimental data after electroforming and the smooth lines are the experimental data of the corresponding devices before electroforming.

The temperature-dependent I-V characteristics of an as-fabricated Nb/NbO$_{2.22}$/Pt device are shown in Figure 4(a) for both positive and negative bias. Comparison of the device current at equivalent positive and negative voltages clearly highlights the asymmetry in contact resistance created by the Schottky barriers at the metal/oxide interfaces. Similar asymmetry was observed for all NbO$_x$ films used in this study, and was found to increase with increasing device area. From the electron affinity of Nb$_2$O$_5$ (3.9 eV) and the work functions of Nb (4.3 eV) and Pt (5.65 eV) the theoretical barrier heights at the Nb/oxide and Pt/oxide interfaces are calculated to be 0.4 eV and 1.75 eV, respectively[34]. As a consequence, the contact resistance is expected to be dominated by the Pt/oxide interface, which is under forward bias when negative voltage is applied to the top electrode.

Experimentally, the current-voltage characteristics were analysed in terms of field-enhanced thermionic emission over a Schottky barrier, as described by[35]:

$$I = AA^*T^2 e^{-(\frac{\Phi_{B0} - \alpha\sqrt{V}}{k_B T})} \qquad .. \qquad .. \qquad (1)$$

where $k_B$ is the Boltzmann constant, $T$ is the absolute temperature, $\Phi_{B0}$ is the zero bias potential barrier height at metal/oxide interface, $\alpha$ is the barrier lowering factor, $A^*$ is the Richardson constant and $A$ is the device area. For both bias polarities the experimental data for a given electric field (V) exhibits a linear trend on a ln(I/T$^2$) vs 1000/T plot as shown in Figures 4(b-c). The slope of these curves yields an activation energy corresponding to the apparent potential barrier at the particular applied field (corresponding to V),

$$E = \Phi_{App} = \Phi_{B0} - \alpha\sqrt{V} \qquad .. \qquad (2)$$

The obtained values were found to decrease with increasing the applied bias, as expected from equation (2) and the zero-bias barrier heights were determined to be 0.19 eV and 0.24 eV for negative and positive bias, respectively (Figure 4(d)). These include contributions from the electrodes and the oxide film and are much lower than the theoretical values. Consequently, they serve as a measure of the resistance asymmetry, rather than actual metal/oxide barrier heights.

The resistance asymmetry is represented more clearly by plotting the ratio of currents measured under negative and positive bias as a function of absolute voltage, as shown for the Nb/NbO$_{2.22}$/Pt device in Figure 4(e). This shows that the current ratio (asymmetry) initially increases with increasing voltage before reaching a maximum at ~0.75 V and then decreases at higher voltages before saturating at a value between 1 and 2. This behaviour is attributed to the dominance of Schottky emission at low fields and an increasing contribution from Poole-Frenkel conduction at higher fields as previously reported for Nb$_2$O$_5$-based Schottky diodes [36]. Significantly, the high-field asymmetry is maintained for all sub-stoichiometric NbO$_x$ films even after electroforming (Figure 4(f)). This reflects the fact that the Schottky barriers in the device area surrounding the filamentary region remain unaffected and continue to influence the current distribution in the devices.

**Effect of oxide stoichiometry and device scaling.**

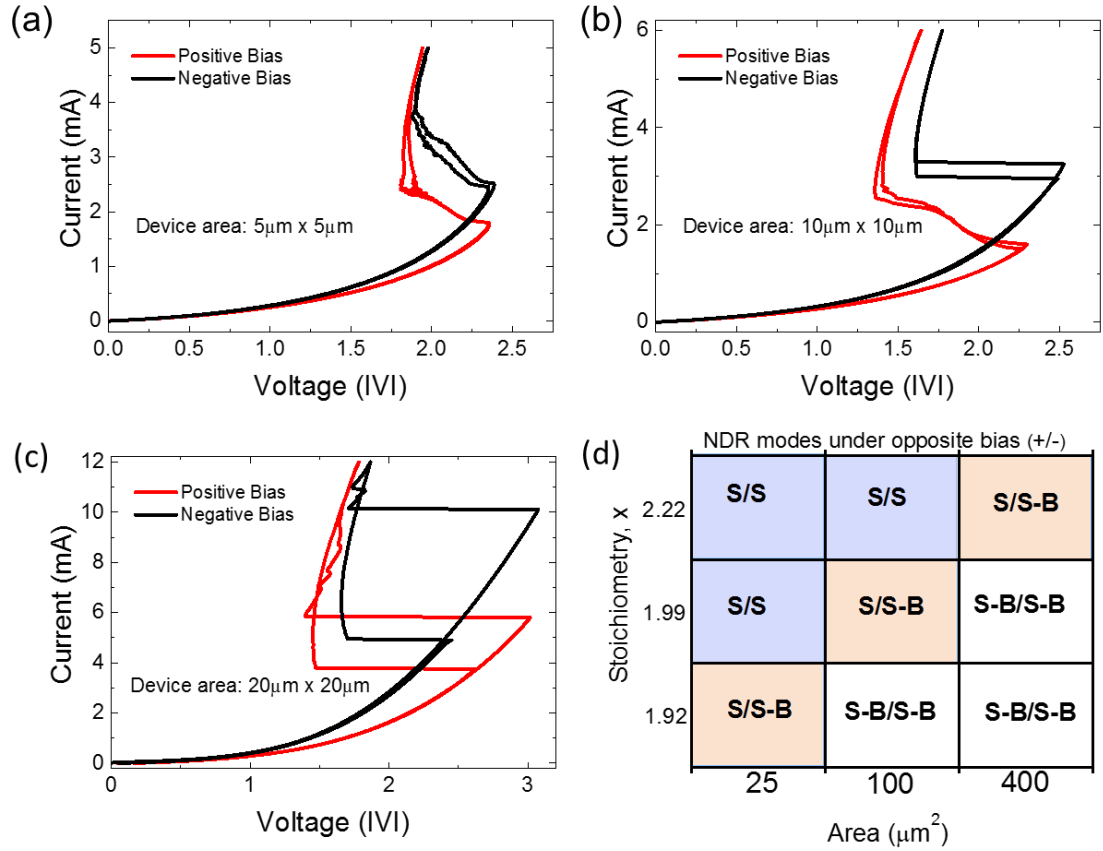

**Figure 5**: (a-c), Asymmetric NDRs observed in Nb/NbO$_{x\sim1.99}$/Pt MOM cross-point devices with 5 μm × 5 μm, 10 μm × 10 μm and 20 μm × 20 μm active area respectively while subsequently biased with positive and negative polarities, and (d) matrix representation of dependency of S-type and snap-back NDR as a function film stoichiometry, device area and bias polarity. The notations S/S and SB/SB are used for S-type NDR and snapback NDR respectively under subsequent opposite biasing conditions, while S/S-B is used to indicate S-type NDR under positive bias and snapback NDR under negative bias.

The film stoichiometry and device area also have a direct impact on the resistance of the region surrounding the electroformed filament and therefore affect the device characteristics. As a consequence, the behaviour reported in Figure 2 is not universal but is specific to devices with particular combinations of device area and film stoichiometry. A systematic study of these dependencies showed that the smallest area devices (2 μm × 2 μm) displayed smooth S-type NDR under both positive and negative biasing regardless of film stoichiometry. In contrast, larger area devices exhibited one of three behaviours depending on the stoichiometry of the oxide film: duel S-type characteristics (Figure 5(a)), combined S-type and snap-back characteristics (Figure 5(b)) or dual snap-back characteristics (Figure 5(c)). These dependencies are summarised in Figure 5(d) which clearly highlights the three distinct regimes. Note that higher sub-threshold and threshold currents were always observed for negative bias polarity, consistent with higher contact resistance from the reverse biased Pt/NbO$_x$ interface.

**Discussion.**

The above data clearly show that the NDR characteristics of simple two-terminal metal/oxide/metal devices are sensitive to the relative resistances of the filamentary conduction path and the surrounding device area, including effects due to polarity dependent contact resistance that arise from the metal/oxide contacts. While the dependencies appear complex, they can be understood with reference to a recently developed parallel-memristor or core-shell model of NDR[25].

In its simplest form the core-shell model assumes that the current distribution in post-formed metal-insulator-metal (MIM) structures is localized, with the NDR response dominated by the high current density filament (core) while the surrounding film (shell) acts as a parallel resistance, as shown schematically in Figure 6. Analysis then shows that the threshold switching response depends on the relative magnitudes of the maximum negative differential resistance $|R_{NDR}|$ of the core and the shell resistances ($R_S$). Specifically, S-type NDR characteristics are observed when $R_S > |R_{NDR}|$ and snap-back characteristics are observed when $R_S < |R_{NDR}|$. Since the magnitude of $R_S$ depends on the conductivity (stoichiometry), area (A), and thickness (t) of the oxide film the model predicts a transition from S-type to snap-back characteristics for critical values of these parameters, as previously demonstrated[30].

For the case where $R_S$ is independent of bias polarity, the above model predicts either smooth S-type or abrupt snap-back characteristics under both positive and negative bias, depending on whether $R_S > |R_{NDR}|$ or $R_S < |R_{NDR}|$. However, when $R_S$ is polarity dependent it is necessary to consider three distinct scenarios. Assuming that $R_S=R_p$ under positive bias and $R_S=R_n$ under negative bias and that $R_p > R_n$ we need to specifically consider the three following cases: a) for $R_p > R_n > R_{NDR}$, S-type characteristics are expected under both positive and negative bias, as shown in Figure 5a; b) for $R_{NDR} > R_p > R_n$, snap-back characteristics are expected under both positive and negative bias, as shown in Figure 5c; and c) for $R_p > R_{NDR} > R_n$, we expect snap-back NDR under negative bias and a smooth S-type NDR under positive bias, as observed in Figure 5(b). The simple criterion defined by the core-shell model therefore provides the basis for understanding the polarity dependent switching characteristics and its dependence on film stoichiometry and device area. In this context it is interesting to note that the current asymmetry is almost negligible in stoichiometric $Nb_2O_5$ as the film resistance in the shell region is much higher than the interface resistance ($R_{film} \gg R_{int.}$). As a consequence these devices showed symmetric S-type NDR as well as symmetric sub-threshold and threshold switching parameters under positive and negative bias.

The effect of film stoichiometry and Schottky barrier heights on NDR characteristics can be illustrated using a lumped element circuit model implemented in LT-SPICE. In this case, the conductivity of the core and surrounding film are assumed to be governed by Poole-Frenkel conduction with the core temperature determined by Joule heating, and that of the surrounding film assumed to remain at ambient temperature. The shell resistance is additionally assumed to include a contribution from the Schottky barrier created by the metal/oxide interfaces, as governed by the Schottky diode equation (equation 1). The shell resistance is therefore given by $R_S = R_{film} + R_{int.}$, as illustrated schematically in Figures 6(a) and 6(b). Using this model, the I-V characteristics of devices were calculated as a function of film and interface resistance, as shown in Figures 6(c) and 6(d), respectively.

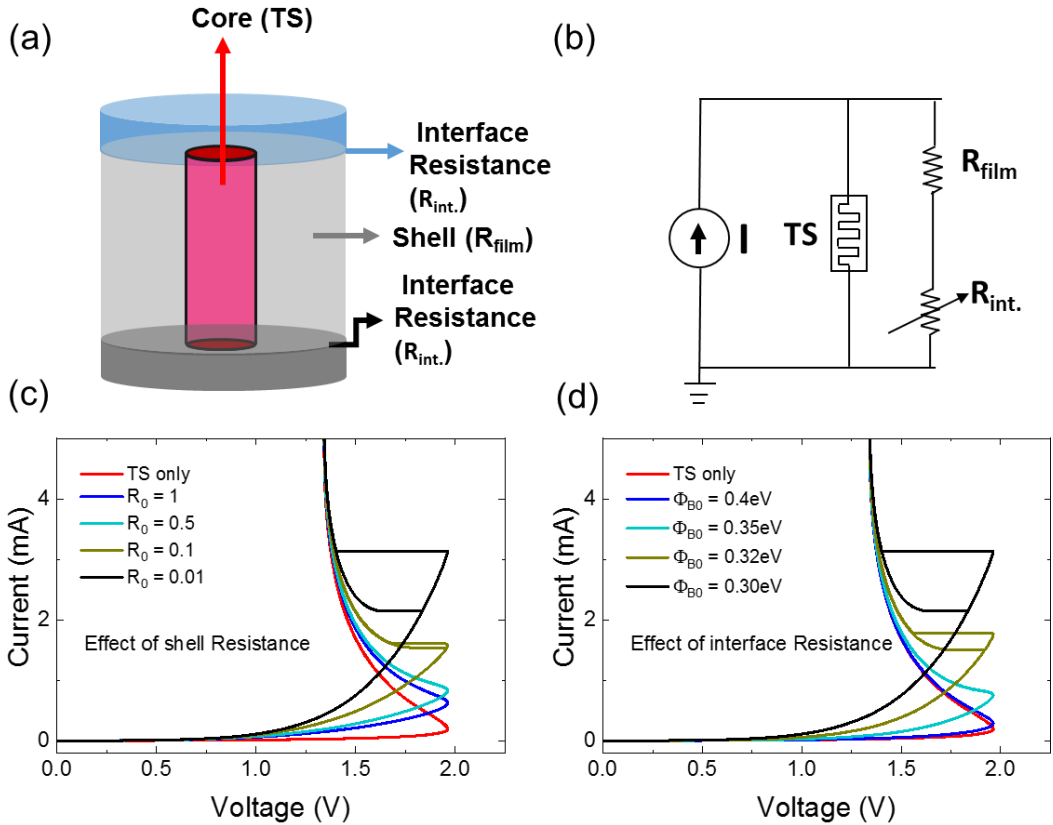

**Figure 6**: (a) Schematic of the core-shell model of memristor, (b) circuit model showing a Threshold switch (TS) as the core element, a variable resistor indicating the interface resistances at the opposite interfaces and $R_{film}$ represents the oxide resistance in the shell area which was modeled by Poole-Frenkel equation, c) effect of shell resistance on the NDR response, (d) effect

of interface resistance (as a function of increasing barrier heights) on the NDR response when the shell resistance ($R_{film}$) is fixed for a given device area, thickness and oxide stoichiometry. Note that, for opposite biasing conditions one interface becomes forward-biased and the other becomes reverse-biased and vice versa, as a result the interface resistance experiences significant change under polarity reversal. Details of the model and the corresponding parameters are given in the supplementary information.

The data clearly demonstrate the transition from smooth S-type to abrupt snap-back NDR characteristics as the shell resistance is reduced, and more detailed analysis confirms that this occurs at $R_{NDR}$. Significantly, a difference in Schottky barrier height of as little as ~0.03 eV is shown to be sufficient to cause such a transition (Figure 6(d)).

**CONCLUSIONS**

In summary, we have demonstrated polarity dependent negative differential resistance characteristics in Nb/NbO$_x$/Pt cross-point devices and shown that the observed asymmetry is a direct consequence of the Schottky barriers created by the metal/oxide contacts. Three distinct behaviours were observed under opposite polarity testing: bipolar S-type NDR, bipolar snap-back NDR, and combined S-type and snap-back NDR, depending on the stoichiometry of the oxide film and device area. By combining ex-situ current-controlled electrical measurements and in-situ thermoreflectance imaging we showed that these NDR responses are associated with strong current localisation within the device. This was used to justify the use of a previously developed two-zone, core shell thermal model of the device. The observed polarity dependent NDR responses, and their dependence on stoichiometry and area were then explained by extending this model to include the effect of the polarity dependent contact resistance. Our extended model provides improved understanding of the NDR response of metal/oxide/metal structures and informs the engineering of devices for neuromorphic computing and non-volatile memory applications.

## AUTHOR INFORMATION

**Corresponding Author**

*E-mail: shimul.nath@anu.edu.au (S. K. Nath)

*E-mail: rob.elliman@anu.edu.au (R. G. Elliman)

## ACKNOWLEDGEMENTS

This work was partly funded by an Australian Research Council (ARC) Linkage Project (LP150100693) and Varian Semiconductor Equipment/ Applied Materials. We would also like to acknowledge the Australian Facility for Advanced ion-implantation Research (AFAiiR), a node of the NCRIS Heavy Ion Accelerator Capability, and the Australian National Fabrication Facility (ANFF) ACT node for access to their research facilities and expertise.

## ASSOCIATED CONTENT

**Supplementary Information.**

### 1. Deposition Conditions of NbOₓ fims

**Table S1**: Film deposition details including the stoichiometry and thickness of each layer.

| Sputter Target | Deposition conditions for NbOₓ layers | | | | | Resisitivity ($\Omega.m$) |
| --- | --- | --- | --- | --- | --- | --- |
| | Power (W) | Pressure (mTorr) | Gas flow (Ar/O$_2$) sccm/sccm | Thickness (nm) | x in NbOₓ | |
| Nb$_2$O$_5$ | 180 (rf) | 4 | 20/0 | 45 | 2.6±0.05 | 3.5x10$^4$±1.5x10$^3$ |
| Nb | 100 (dc) | 1.5 | 19/1 | 44.3 | 1.92±0.04 | 8±1.5 |
| Nb | 100 (dc) | 2 | 19/1 | 55 | 1.99±0.03 | 15±1 |
| Nb | 100 (dc) | 1.5 | 18.5/1.5 | 57.3 | 2.22±0.04 | 120±7 |

### 2. Electroforming and subsequent I-V sweeps of devices with NbO$_{1.92}$ film

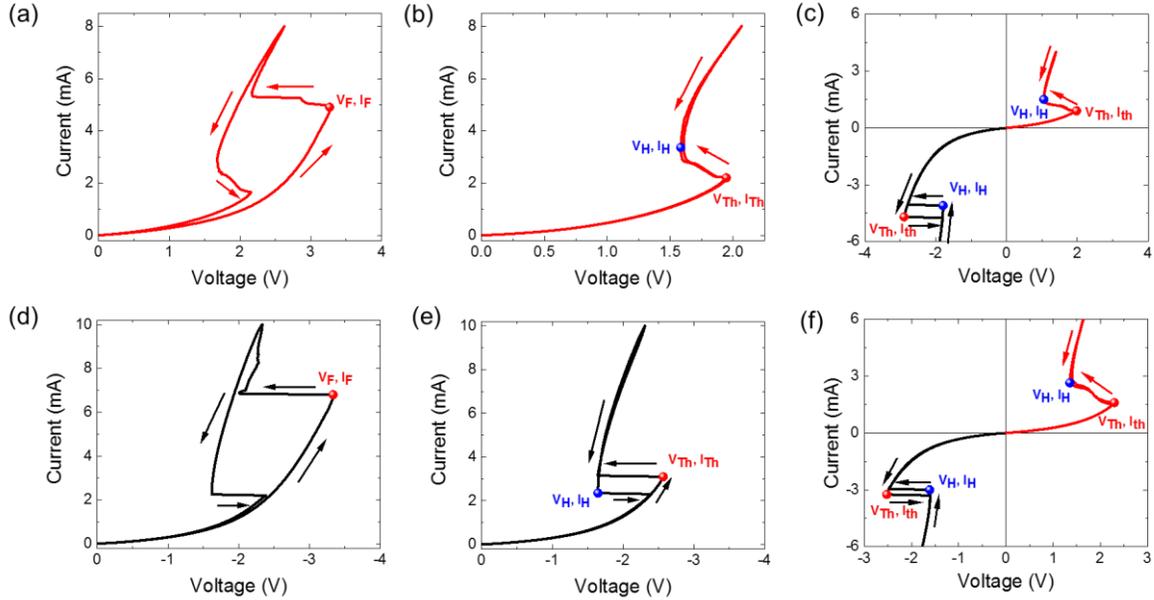

**Figure S1.** Polarity dependent electroforming and CC-NDR characteristics: (a) electroforming and (b) subsequent switching characteristics of a 5 µm × 5 µm cross-point devices under positive

bias, (c) transition of S-type into snapback when negative bias is applied to the top electrode, (d) electroforming and (e) subsequent switching characteristics of a 5 μm × 5 μm cross-point devices under negative bias, and (f) transition of snap-back into S-type NDR when positive bias is applied to the top electrode.

## 3. SPICE Models for archetype threshold switching memristor with Poole-Frenkel model

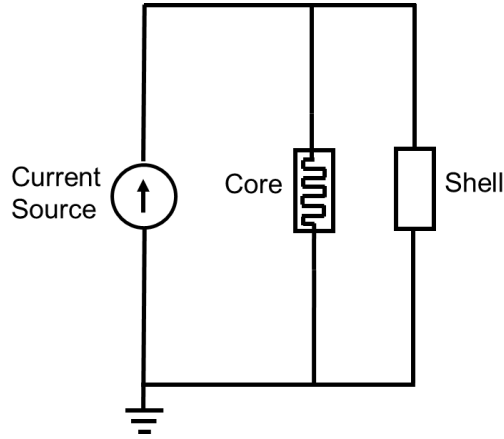

**Figure S2.** SPICE equivalent circuit of a core-shell structure for one active memristor and a shell region in parallel.

We considered an archetype threshold switch (memristor) in our model as the core element with a series combination of a parallel resistor (representing the film resistance in the shell region) and a constant Schottky barrier (represented by equation (1) which represents the interface barriers at the metal/NbO$_x$ interfaces). The electrical conductivity of the core region was assumed to be Poole-Frenkel emission such that the device resistance is given by[7]:

$$R_m = R_0 e^{\frac{1}{k_B T}(E_a - q\sqrt{\frac{qE}{\pi\varepsilon_0\varepsilon_r}})} \quad\quad\quad (3)$$

where $k_B$ is the Boltzmann constant, $E_a$ is the activation energy, $\varepsilon_0$ is the vacuum permittivity, and $\varepsilon_r$ is the relative permittivity of the threshold switching volume and surrounding NbOx film. $T_m$ and $T_{amb}$ denote the temperature of the electrically active region and the ambient environment, $R_0$ is the resistance pre-factor of the active region at $T = T_{amb}$.

The dynamic behaviour of the memristor is defined by Newton's law of cooling:

$$\frac{dT_m}{dt} = \frac{I_m^2 R_m}{C_{th}} - \frac{\Delta T}{R_{th} C_{th}} \quad \quad \quad (4)$$

where $R_{th}$ and $C_{th}$ are the thermal resistance and the thermal capacitance of the device, and $\Delta T$ is the temperature difference between the $T_m$ and $T_{amb}$.

The parallel resistor, $R_{film}$ was also modelled by a similar Poole-Frenkel equation (eq. 3) with identical activation energy, but without any thermal feedback. The Schottky barrier represented by equation (1) was series connected to $R_{film}$. Details of the model parameters are given in the Table-S2.

**Table S2.** Memristor parameters used in simulation for core memristor and shell resistance in the Poole-Frenkel model.

| Model Parameters (Unit) | Symbol | Threshold Switch (Core) | Shell Resistance, $R_{film}$ | Schottky parameters |
|---|---|---|---|---|
| Thermal capacitance (J·K⁻¹) | $C_{th}$ | $1 \times 10^{-15}$ | | |
| Resistance prefactor (Ω) | $R_0$ | 80 | 0.01 to 1 | |
| Thermal resistance (K·W⁻¹) | $R_{th}$ | $2 \times 10^5$ | $2 \times 10^5$ | |
| Ambient temperature (K) | $T_{amb}$ | 298 | 298 | |
| Metallic state resistance (Ω) | $R_v$ | 0 | 0 | |
| Activation Energy (eV) | $E_a$ | 0.23 | 0.23 | |
| Boltzmann constant (J·K⁻¹) | $k_B$ | $1.38 \times 10^{-23}$ | $1.38 \times 10^{-23}$ | |
| Elementary charge (C) | e | $1.6 \times 10^{-19}$ | $1.6 \times 10^{-19}$ | |
| Vacuum permittivity (F·m⁻¹) | $\varepsilon_0$ | $8.85 \times 10^{-12}$ | $8.85 \times 10^{-12}$ | |

| Relative permittivity of the threshold switching volume and surrounding NbOx film | $\varepsilon_r$ | 45 | 45 | |
| --- | --- | --- | --- | --- |
| Richardson's constant (A·K$^{-2}$·m$^{-2}$) | $A^*$ | | | 480 |
| Barrier lowering factor (eV) | $\alpha$ | | | 0.24 |
| Zero bias barrier (eV) | $\Phi_{B0}$ | | | 0.3 to 1 |
| Schottky diode area (m$^2$) | $A$ | | | $2.48 \times 10^{-11}$ |
| Film thickness (nm) | | 45 | | |

## 4. Asymmetric threshold currents for different area in sub-stoichiometric NbO$_x$

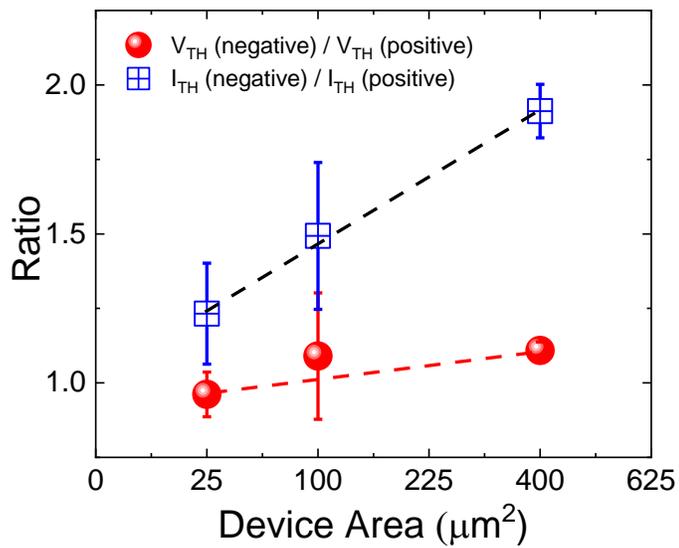

**Figure S3.** Sub-stoichiometric NbO$_x$ (x=1.92) with three distinct device area showing asymmetric threshold parameters.

## 5. Voltage-controlled threshold switching in sub-stoichiometric NbO$_x$ and stoichiometric Nb$_2$O$_5$.

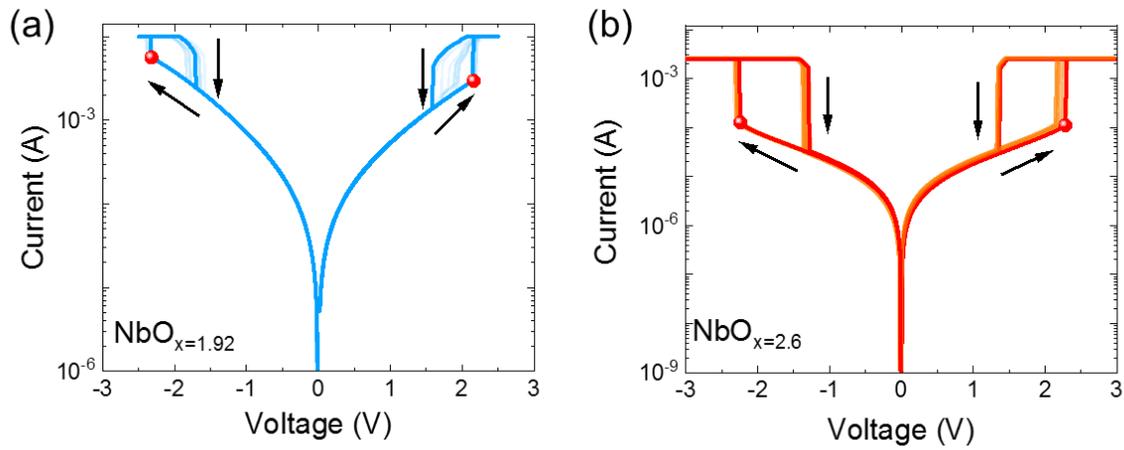

**Figure S4.** (a-b) Asymmetric and symmetric threshold switching in two identical 20 μm × 20 μm devices with sub-stoichometric NbO$_x$ (x=1.92) and nearly-stoichiometric Nb$_2$O$_5$ (x=2.6) respectively under voltage-controlled operation.